\documentclass[useAMS,usenatbib,usegraphicx]{mn2e}

\def\Msun{M_\odot}
\newcommand\brittion[2]{\mbox{#1\hspace{0.2em}{\rmfamily{#2}}}} 

\def\aj{AJ}%
\def\apj{ApJ}%
\def\apjl{ApJ}%
\def\apjs{ApJS}%
\def\aap{A\&A}%
\def\mnras{MNRAS}%
\def\procspie{Proc.~SPIE}%

\title[The Dark Matter Haloes and Host Galaxies of \brittion{Mg}{II} Absorbers at z$\sim$1]{The Dark Matter Haloes and Host Galaxies of \brittion{Mg}{II} Absorbers at z$\sim$1}
\author[Lundgren, Wake \& Padmanabhan et al.]{ \parbox{\textwidth}{
Britt F. Lundgren$^{1}$,
David A. Wake$^{1}$,
Nikhil Padmanabhan$^{1}$, 
Alison Coil$^{2,3}$, 
Donald G. York$^{4,5}$.}\\\\
$^{1}$Yale Center of Astronomy and Astrophysics, Yale University, New Haven, CT 06511, USA\\
$^{2}$Department of Physics, University of California at San Diego, San Diego, CA, USA\\
$^{3}$Alfred P. Sloan Foundation Fellow\\
$^{4}$Department of Astronomy and Astrophysics, University of Chicago, Chicago, IL 60637, USA\\
$^{5}$Enrico Fermi Institute, 5640 Ellis Ave., Chicago, IL 60637, USA\\}

\begin{document}

\date{}

\pagerange{\pageref{firstpage}--\pageref{lastpage}} \pubyear{2011}

\maketitle

\label{firstpage}

\begin{abstract}
Strong foreground absorption features from singly-ionized Magnesium (\brittion{Mg}{II}) are commonly observed in the spectra of quasars and are presumed to probe a wide range of galactic environments.  To date, measurements of the average dark matter halo masses of intervening \brittion{Mg}{II} absorbers by way of large-scale cross-correlations with luminous galaxies have been limited to z$<$0.7.  In this work we cross-correlate 21 strong (W$_{r}^{\lambda2796}$$\ga$0.6\AA) \brittion{Mg}{II} absorption systems detected in quasar spectra from the Sloan Digital Sky Survey Data Release 7 with $\sim$32,000 spectroscopically confirmed galaxies at 0.7$\leq$z$\leq$1.45 from the DEEP2 galaxy redshift survey.  We measure dark matter (DM) halo biases of $b_{G}$=1.44$\pm$0.02 and $b_{A}$=1.49$\pm$0.45 for the DEEP2 galaxies and  \brittion{Mg}{II} absorbers, respectively, indicating that their clustering amplitudes are roughly consistent.  Haloes with the bias we measure for the \brittion{Mg}{II} absorbers have a corresponding mass of 1.8$\pm^{4.2}_{1.6}\times10^{12}h^{-1}\Msun$, although the actual mean absorber halo mass will depend on the precise distribution of absorbers within DM haloes. This mass estimate is consistent with observations at z=0.6, suggesting that the halo masses of typical \brittion{Mg}{II} absorbers do not significantly evolve from z$\sim$1.  We additionally measure the average W$_{r}^{\lambda2796}$$\geq$0.6\AA~ gas covering fraction to be $f_{c}$=0.5 within 60 h$^{-1}$kpc around the DEEP2 galaxies, and we find an absence of coincident strong \brittion{Mg}{II} absorption beyond a projected separation of $\sim$40 h$^{-1}$kpc. Although the star-forming z$>$1 DEEP2 galaxies are known to exhibit ubiquitous blueshifted \brittion{Mg}{II} absorption, we find no direct evidence in our small sample linking W$_{r}^{\lambda2796}$$\geq$0.6\AA~  absorbers to galaxies with ongoing star formation.
\end{abstract}


\section{Introduction}

Absorption features in quasar spectra (quasar absorption lines; QALs) provide nearly unbiased probes of the gas and dust content of foreground galaxies and the intergalactic medium (IGM) over vast ranges in redshift.  As such, the growing catalogues of QALs now available from large spectroscopic surveys (e.g., SDSS; York et al. 2000) may be used as reliable tracers of the metal abundances of galaxy environments throughout $\sim$90\% of cosmic history.  However, critical questions regarding the physical origins of QALs remain to be resolved.  Metal-rich QALs have a well-established association with $\sim$L$^{*}$ galaxies \citep[e.g.,][]{Bergeron86,Yanny87,Cristiani87, BB91,LB93, Steidel93, Steideletal94}, but whether the absorbing gas primarily traces the cool extended regions of dark matter haloes, dwarf satellite galaxies, galactic disks, or supernovae-driven outflows remains an active matter of debate. 

The root of the ambiguity surrounding the origins of QALs lies in the fact that while these absorbers are easily detected at moderate and high redshifts, where the most prolific UV metal lines (e.g., singly-ionised Magnesium; \brittion{Mg}{II}, a tracer of photoionised cold gas) are redshifted into the optical, their galactic hosts are nearly always too faint to observe.   For a limited number of objects, deep imaging has been used to identify candidate galaxies as hosts of \brittion{Mg}{II} absorbers \citep[e.g.,][]{LB93,Steidel97,Nestor07, B07, Straka10, Chun10, Meiring11}.  The results of these investigations suggest a correlation between absorber equivalent width and host star-formation rate, though many galaxy hosts of large equivalent width absorbers remain undetected in even the deepest imaging.

Recent efforts to determine the typical physical properties of galaxies hosting strong\footnote[1]{For brevity, throughout this work we will refer to strong absorbers as those with W$_{r}^{\lambda2796}$$\ga$0.6\AA, where W$_{r}^{\lambda2796}$ is the equivalent width of the 2796\AA~ transition of \brittion{Mg}{II} in the absorber rest frame.} \brittion{Mg}{II} absorbers have attempted statistical analyses using large absorber catalogues.  The typical luminosities and colors of \brittion{Mg}{II} hosts have been inferred from the stacking of residual light around thousands of quasars exhibiting \brittion{Mg}{II} absorption \citep{Zibetti07, Caler10}. Host star-formation rates and dust properties have been inferred from spectral stacks of hundreds of identified absorbers \citep{York06,Wild07, Menard09, MC09, Menard10, ND10}, and average absorber dark matter halo masses have been indirectly obtained from \brittion{Mg}{II} clustering measurements \citep{B06, L09, Gauthier09}. The imaging and spectral stacking analyses collectively indicate a general correlation between \brittion{Mg}{II} equivalent width and the star formation rate of the host galaxy, whilst the clustering analyses suggest a weak global anti-correlation of the absorber equivalent width with galaxy halo mass. Taken together, these findings disfavor the classical picture in which strong \brittion{Mg}{II} absorption is primarily produced by virialised gas in the extended haloes of massive galaxies.  

Now a new paradigm seems to be emerging in which the majority of strong \brittion{Mg}{II} absorbers originate in galactic disks on scales $<$10 h$^{-1}$ kpc and in star formation-driven outflows on larger scales, out to $\sim$150 kpc.  Deep imaging and spectroscopic follow-up of galaxy-absorber pairs at z$>$0.7  have confirmed the existence of a strong correlation between W$^{\lambda2796}_{r}\ga$2\AA~ absorbers and starburst galaxies \citep{B07,Nestor10}.  Further strengthening the case for supernovae-driven outflows is the recent determination that the redshift number density evolution of strong \brittion{Mg}{II} absorbers traces the global star formation rate \citep{Menard09,CB10}. 

However, studies of \brittion{Mg}{II} absorbers in the local Universe have shown that the origins are less clear in the case of the much larger population of absorbers with W$^{\lambda2796}_{r}\la$2\AA.  In an examination of such lower equivalent width \brittion{Mg}{II} absorbers at z$\sim$0.2, \citet{Chen10} found no compelling evidence of a correlation between absorption properties and the recent star formation histories of galaxies.  Furthermore, recent studies of the galactic counterparts of W$^{\lambda2796}_{r}\la$2\AA~absorbers at z$\sim$0.1 find no evidence for outflows from star formation \citep{K11}, suggesting that \brittion{Mg}{II} absorption primarily probes infalling gas from disk/halo processes in normal galaxies, at least for absorbers with similar equivalent widths in the local Universe.  To what extent galaxy evolution or galaxy sample selection biases may be able to reconcile the contradictory findings of \brittion{Mg}{II} origins in star formation-driven outflows at z$\ga$0.5 and cold mode accretion at z$\la$0.3 still remains to be determined.

For large data sets, clustering measurements can be used to determine the typical environments of any well-defined sample of extragalactic objects.  The space density of strong \brittion{Mg}{II} absorbers is low -- on average, only 0.3 absorbers with W$^{\lambda2796}_{r}\ga$0.6\AA~ are observed per unit redshift along quasar sightlines at z=0.5  \citep{Nestor05} -- so \brittion{Mg}{II} clustering estimates in the aforementioned works utilised two-point cross-correlations with much more numerous luminous galaxies in order to extract an absorber clustering signal of measurable strength. Luminous red galaxies (LRGs) have thus far been a favored galaxy sample for this purpose, since their observable redshift range in the SDSS has a large overlap with \brittion{Mg}{II} absorbers.  In addition to their convenient redshift range, LRGs are excellent tracers of structure in the Universe \citep[e.g.,][]{Eisenstein05a, Zehavi05, Ross07, Blake08, Wake08a}, and they exhibit minimal stellar mass evolution over their observable redshift range in the SDSS (z$\la$0.8) \citep{Wake06, Wake08a, Brown07, Brown08, Cool08}.  The dark matter halo bias of LRGs is thus precisely measured \citep{Blake08, Wake08a, NP09}, making these galaxies a favorable population for  amplifying the clustering signals of quasars and absorbers.   Measurements of the \brittion{Mg}{II}-LRG two-point cross-correlation have converged on a typical \brittion{Mg}{II} dark matter halo mass of $\sim$10$^{12}$ M$_{\sun}$ at z$\sim$0.6 \citep{B04, B06, L09, Gauthier09}.   


Few deep spectroscopic surveys exist with sufficient numbers of galaxies to attempt measurements of large-scale absorber clustering (and thus, the dark matter bias) at higher redshifts (z$>$0.6).  Lyman break galaxies (LBGs) have been used to examine the bias of strong quasar absorption lines at z$>$2; \citet{Adelberger05} measured the clustering of triply-ionised Carbon (\brittion{C}{IV}) absorption lines ($\lambda\lambda$ 1548,1551) around UV-selected galaxies at 2$<$z$<$3  finding that \brittion{C}{IV} absorbers cluster similarly to typical star-forming galaxies.  Still, a similar analysis has not yet been performed for a sample of \brittion{Mg}{II} absorbers, which trace gas with a lower ionization temperature and are thus not always coincident with \brittion{C}{IV} absorption.  

The DEEP2 Galaxy Redshift Survey \citep{Davis03} provides a sample of $\sim$32,000 galaxies with spectroscopically confirmed redshifts in the range 0.7$<$z$<$1.45, ideal for use in correlations with \brittion{Mg}{II} detected in the SDSS.  This galaxy sample is particularly interesting for \brittion{Mg}{II} investigations, as its star-forming population has been recently found to exhibit ubiquitous \brittion{Mg}{II} outflows (observed as blueshifted intrinsic absorption in the stacked rest-frame galaxy spectra) across a wide range of galaxy properties \citep{Weiner09}.  The outflows reported for the DEEP2 galaxies have typical column densities of $N_{H}\sim10^{20}$ cm$^{-2}$ and velocities of $\sim$400 km s$^{-1}$.  These values are consistent with observations from quasar absorption line data, and \citet{Weiner09} speculate that the ejected gas could eventually reach projected separations of 50 h$^{-1}$kpc.  However, no direct connection between feedback from normal star-forming galaxies and quasar absorption lines has yet been observed.

In this work, we calculate the two-point redshift-space cross-correlation of  the full sample of DEEP2 galaxies with 21 \brittion{Mg}{II} absorbers identified in SDSS DR7 quasar spectra within the DEEP2 survey area.  By comparing to the average dark matter halo bias of the galaxy sample, obtained through a calculation of the galaxy auto-correlation, we estimate the typical bias of the absorbers and thereby quantify their average halo mass.  We also measure the covering fraction of W$^{\lambda2796}_{r}\ga$0.6\AA~ absorption for the DEEP2 galaxy sample and compare with previous results.  A description of the data is given in Section 2.  The methodology we apply for the analysis is described in Section 3, and results are presented in Sections 4-7.  A discussion of the results is given in Section 8, and a summary of important findings is presented in Section 9.

Throughout this paper, we assume a flat $\Lambda$--dominated CDM cosmology with $\Omega_m=0.27$, $H_0=73$ km s$^{-1} $Mpc$^{-1}$, and $\sigma_8=0.8$ unless otherwise stated.

\section{Data}

\subsection{The \brittion{Mg}{II} Absorber Sample}

The quasar spectra used in this analysis are drawn from the Sloan Digital Sky Survey Seventh Data Release \citep[SDSS DR7][]{A09}.  The SDSS DR7 Quasar catalogue \citep{Schneider10} contains 88 quasars in regions of overlap with the DEEP2 survey.  Each of the quasars in the DEEP2 survey area was run through an automated pipeline that detects strong \brittion{Mg}{II} absorption systems with high precision (for full description, see York et al. 2011). In brief, this pipeline first searches for absorption features in quasar spectra.  For each detected absorption feature, a Gaussian profile is fit to the normalised flux to extract precise centroid and equivalent width measurements.  In order to identify the ion and redshift corresponding to each measured absorption line, a system-finding algorithm then identifies pairs of 4$\sigma$ absorber detections matching the wavelength separation expected for \brittion{Mg}{II} at a given redshift.  The reliability of the identification is further quantified for each detected absorption system, taking into account the doublet ratio measured for \brittion{Mg}{II}, the number of additional ions matched in absorption at the same redshift, and any blending with other absorption features identified at another redshift.  


Due to the magnitude-limited design of the SDSS survey, the spectral signal-to-noise ratio (SNR) correlates with the optical apparent magnitude of each quasar.  The completeness of the SDSS absorption line data for any object thereby varies simultaneously with absorber equivalent width and quasar magnitude, such that the fainter the quasar, the poorer the spectral SNR, and thus the larger the minimum absorber equivalent width for detection at the required 4$\sigma$ limit.  A full description of the detection completeness is provided in \citet{York11}, but generally z$\sim$1 \brittion{Mg}{II} absorption systems with W$_{r}^{\lambda2796}$$>$0.6\AA~ are detected with $>$90\% completeness in SDSS quasars with m$_{i}\leq$20.  Because we include in our sample quasars with m$_{i}>$20, all candidate absorption systems were verified by visual inspection to rule out any false detections due to artifacts such as problematic continuum subtraction around narrow emission lines or noise spikes.  

In all, we find 21 \brittion{Mg}{II} absorption systems with W$_{r}^{\lambda2796}$$\geq$0.6\AA~ and a velocity separation of $v>$15,000 km s$^{-1}$ in the quasar rest frame.  This velocity difference is generally sufficient to ensure that the absorbers are unambiguously unassociated with the background quasar and thus originating in foreground galactic environments \citep{Wild08}.  The mean rest-frame equivalent width of the 18 absorbers extracted from spectra with m$_{i}\leq$20 is $<$$W_{r}^{\lambda2796}$$>$=1.19\AA~, compared to $<$$W_{r}^{\lambda2796}$$>$=1.88\AA~ for the remaining 3 absorbers drawn from fainter quasars.  While we expect that the three included spectra with m$_{i}>$20 are only complete to W$_{r}^{\lambda2796}$=1.0\AA, these data represent a small fraction of the selected sample.  Additional details of the \brittion{Mg}{II} absorber detections are provided in Table 1.

\begin{figure}
\includegraphics[width=84mm]{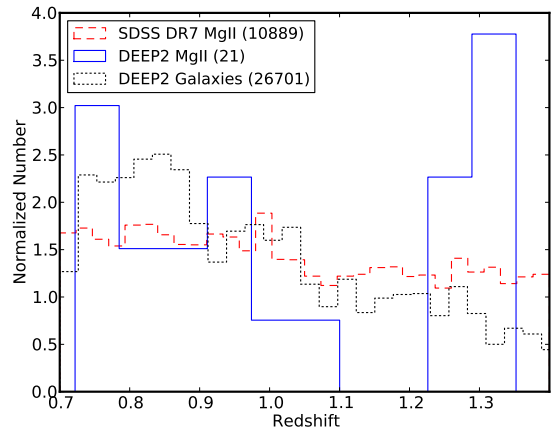}
\caption{The normalised redshift distributions of the 21 \brittion{Mg}{II} absorbers and 26,701 DEEP2 galaxies examined in this work.  We also include the redshift distribution of all $\sim$11,000 \brittion{Mg}{II} absorbers detected over the redshift range of interest in the SDSS DR7 \citep{York11}, indicating that our data is consistent with a random sample of the global population.  The effects of differences in these redshift distributions on the calculation of the \brittion{Mg}{II}-galaxy cross-correlation have been measured and are discussed in Section 4.}
\label{fig:1}
\end{figure}


\subsection{The DEEP2 Galaxy Sample}

The galaxies used in this analysis are taken from the DEEP2 Galaxy Redshift Survey \citep{Davis03}, which obtained spectra for $\sim$32,000 galaxies using the DEIMOS spectrograph \citep{Faber03} on the Keck II telescope.  The DEEP2 survey observed galaxies in the redshift range 0.7$<$z$<$1.45, with a limiting magnitude of $R_{AB}$=24.1.  A magnitude cut of $18.5<R_{AB}<24.1$, in combination with a BRI color cut,  was applied to select galaxies in the redshift range of interest from CFHT CFH12K imaging \citep{Coil04b}.   The survey geometry consists of four spatially separated fields, which together cover 3 deg$^{2}$ of the sky.  Additional details of the survey observations and data reduction are available in \citet{Davis03}, and \citet{Coil04a}.   The DEEP2 galaxy catalog is a flux-limited sample with $<m_{R}>$=23.3 and a mean rest-frame B-band luminosity $<M_{B}>$=-20.1, which has been used extensively for clustering measurements at z$\sim$1.  For this work, we apply the exact sample used by \citet{Coil07} to measure the quasar-galaxy cross-correlation.

\section{The Two-Point Cross-Correlation Function}

Due to the extremely low space density of the \brittion{Mg}{II} systems in our sample, we are unable to measure the absorber clustering by way of an auto-correlation function.  However, we can exploit the high space density of the DEEP2 galaxies to measure the clustering of the \brittion{Mg}{II} systems via their cross-correlation with these galaxies.

The two-point cross-correlation function between two sets of objects {\it a} and {\it b}, $\xi_{ab}(s)$, 
is defined as a measurement of the excess probability above Poisson of finding an object {\it a} 
at a separation {\it s} from another object {\it b}. In order to calculate the \brittion{Mg}{II}-galaxy two-point correlation function, we compare the number of \brittion{Mg}{II}-galaxy pairs to the number of \brittion{Mg}{II}-random pairs as a function of scale such that,

\begin{equation}
\label{eq:cross}
	\xi(s) = \frac{n_R}{n_G}\frac{N_{AG}(s)}{N_{AR}(s)} - 1,
\label{eq:agcross}
\end{equation}
where $N_{AG}$ and $N_{AR}$ are the absorber-galaxy and absorber-random pair counts respectively, 
$n_G$ and $n_R$ are the number of DEEP2 galaxies and DEEP2 random galaxies respectively, and $s$ is the pair separation in redshift space. We choose to use this estimator for the cross-correlation since it only requires a random catalogue for the galaxies, and hence only knowledge of the galaxy selection function is required. This greatly simplifies our calculation, since the selection function of the \brittion{Mg}{II} absorbers is complex.

The DEEP2 random galaxy catalogue is generated following the spatially dependent selection function of the DEEP2 galaxies, which takes into account the window function of the survey and the spatially dependent redshift success rate, along with the redshift selection function. Details of this procedure are given in \citet{Coil07}. We generate a random catalogue that is 1000 times larger than the galaxy catalogue, which is sufficient to ensure shot noise in the absorber-random pair counts does not contribute significantly to our error budget.

We generate errors on the cross-correlation by bootstrap resampling the absorber systems. We draw 1000 randomly selected samples of 21 absorbers including repeats and recalculate the cross-correlation function. We then find the central 68\% of these measurements to define the 1$\sigma$ error on each point. Normally such a technique would be unsuitable for a clustering measurement, but the absorbers are so spatially sparse that they are essentially independent of each other.

Since the cross-correlation function is related to the auto-correlation functions of both input samples, we need to measure the auto-correlation of the the DEEP2 galaxies in order to determine the clustering properties of the \brittion{Mg}{II} absorbers.  For this measurement we use the \citet{Landy93} estimator as:
\begin{equation}
\label{eq:cross2}
	\xi(s) = \frac{N_{GG}(s) - 2N_{GR}(s)+N_{RR}(s)}{N_{RR}(s)} - 1,
\end{equation}
where $N_{GG}$, $N_{GR}$, and $N_{RR}$ respectively represent the galaxy-galaxy, galaxy-random, and random-random pair counts.  This  slightly more complex estimator differs from what was used to calculate the absorber-galaxy cross-correlation;  however, we repeated the calculation of the galaxy auto-correlation using an estimator symmetric to Eq. \ref{eq:agcross} and found no significant difference in the results. 

We use the same randoms as before but this time estimate the errors using jackknife resampling, removing each of the 10 DEEP2 DEIMOS fields, one at a time. Whilst one would normally need more than 10 fields to accurately estimate the errors and covariance of a clustering measurement using this technique, our overall error budget is dominated by the errors on the cross-correlation.  We thereby find this small number of fields to be sufficient for our purposes.   

\begin{figure}
\centering
\includegraphics[width=84mm]{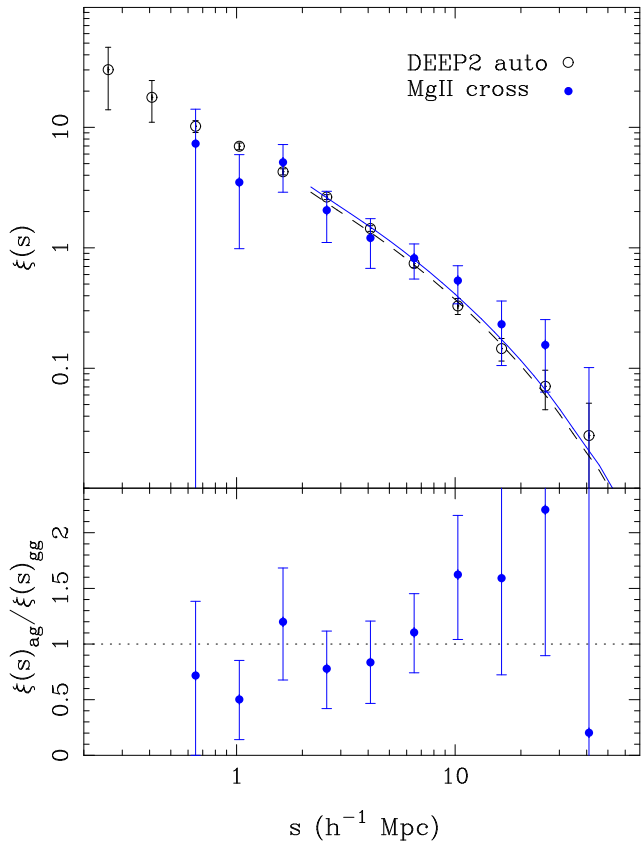}
\caption{The two-point redshift-space cross-correlation of \brittion{Mg}{II} absorption line systems and DEEP2 galaxies is shown in comparison with the auto-correlation of the same DEEP2 galaxies, each with 1$\sigma$ confidence intervals overplotted. The galaxy auto-correlation and absorber-galaxy cross-correlation are consistent within each other on all scales suggesting that the absorbers occupy the same average environments as the DEEP2 galaxies.}
\label{fig:2}
\end{figure}

We show in Figure \ref{fig:2} the redshift space \brittion{Mg}{II} absorber-DEEP2 galaxy two-point cross-correlation function (filled circles) compared to the two point auto-correlation of the galaxies (open circles). Within the errors the \brittion{Mg}{II}-galaxy cross-correlation and the galaxy auto-correlation are indistinguishable, implying that on scales larger than 0.6 $h^{-1}$ Mpc \brittion{Mg}{II} absorbers live in the same environments as the DEEP2 galaxies.

\section{The Bias and Dark Matter Halo masses of \brittion{Mg}{II} absorbers}

The relationship between the clustering of galaxies and the clustering of the underlying dark matter distribution is often described by the bias ($b$) parameter. In the simplest case the bias ($b$) is just assumed to be independent of scale so that $\xi_{gal}(r) = b^2\xi_{DM}(r)$ and so it is straightforward to estimate $b$ if $\xi_{DM}(r)$ is known.

In our case we have measured the redshift space correlation function, $\xi(s)$, rather than the real space correlation function, $\xi(r)$, and so must take into account the effect that the peculiar velocities of galaxies have on the distance measurements in the redshift direction. There are two terms that contribute to the peculiar velocity: the small scale random motion of galaxies with DM haloes and the large scale coherent streaming of galaxies from underdense to overdense regions, known as the Kaiser effect \citep{Kaiser87}. Since we only wish to measure the bias we can restrict ourselves to consider only the large scale clustering amplitude which means we only include the Kaiser effect when converting from $\xi(r)$ to $\xi(s)$.

\citet{Kaiser87} showed that one can relate $\xi(s)$ to $\xi(r)$ as

\begin{equation}
	\xi(s) = \xi(r)\Bigl(1 + \frac{2}{3}\beta + \frac{1}{5}\beta^2\Bigr),
\label{eq:kaia}
\end{equation}

 where 

\begin{equation}
	\beta \simeq \frac{\Omega_m^{0.55}}{b}
\end{equation}
and $\Omega_m$ is the matter density of universe at the redshift of interest.

On large scales the cross-correlation can be determined from the respective auto-correlations with $\xi_{AG}^2 = \xi_A\xi_G$, and it is related to the dark matter clustering as $\xi_{AG}(r) = b_Ab_G\xi_{DM}(r)$. Therefore the conversion from real space to redshift space must include $\beta$ terms for both the absorbers and the galaxies and may be parameterised as

\begin{equation}
	\xi_{AG}(s) = \xi_{AG}(r)\Bigl[1 + \frac{1}{3}(\beta_A + \beta_G) + \frac{1}{5}\beta_A\beta_G\Bigr]
\label{eq:kaic}
\end{equation}
where $\beta_G$ and $\beta_A$ depend on the bias of the galaxies and absorbers, respectively.

We calculate $\xi_{DM}(r)$ at $z = 0.95$, the mean galaxy redshift, as described in \citet{Wake11} using the non-linear power-spectrum produced by the CAMB package\footnote[2]{http://camb.info} which makes use of the HALOFIT routine from \citet{Smith03}. We then measure the DEEP2 galaxy bias by fitting to the galaxy auto-correlation function. Fitting on large scales allows us to include the value of the galaxy bias in Equation \ref{eq:kaic} and fit to the \brittion{Mg}{II}-DEEP2 cross-correlation to extract the absorber bias. Estimates of the scales at which $\xi(s)/\xi(r)$ reach the linear theory asymptote (i.e. there is no effect from peculiar velocities and only the \citet{Kaiser87} corrections need to be applied) range from 4-8 $h^{-1}$ Mpc \citep{Hawkins03,Tinker06}. We choose to make two fits on scales $> 2 h^{-1}$ Mpc and $> 6 h^{-1}$ Mpc. The smaller scale yields a higher signal-to-noise measurement but may be biased due to any difference in the peculiar velocities of absorbers and galaxies resulting in a difference in the suppression of $\xi(s)$. Any such bias is likely to be dwarfed by our measurement errors and indeed both fits are equivalent within the errors. It would of course be possible to overcome these potential biases by using the projected correlation function, but with such low numbers of absorbers the measurement becomes incredibly noisy and essentially unusable.

In order to estimate the errors on the measured cross-correlation we fit to each of our bootstrap samples and again find the central 68$^{th}$ percentile to represent the 1$\sigma$ error. This yields bias measurements for the DEEP2 galaxies of 1.44$\pm$0.02 ($s > 2h^{-1}$ Mpc) and  1.38$\pm$0.05 ($s > 6h^{-1}$ Mpc) and bias measurements of 1.49 $\pm$ 0.45 ($s > 2h^{-1}$ Mpc) and  1.92 $\pm$ 0.61 ($s > 6h^{-1}$ Mpc) for the \brittion{Mg}{II} absorbers. We note that not incorporating the Kaiser effect in our modeling of the measurements produces bias values that are 20\% higher.

We can use these bias measurements to estimate the halo mass of both the absorbers and galaxies. The simplest approach is to determine which mass halos have the same bias as the absorbers or galaxies using the halo mass bias relation of \citet{Tinker10}. By this method, we calculate mean halo masses of  $<logM>=12.19\pm0.03\Msun$ ($s > 2h^{-1}$ Mpc)  and $<logM>=12.09\pm0.08\Msun$ ($s > 6h^{-1}$ Mpc) for the DEEP2 galaxies and $<logM>=12.25\pm^{0.52}_{0.95}\Msun$  ($s > 2h^{-1}$ Mpc) and $<logM>=12.74\pm^{0.46}_{0.71}\Msun$ ($s > 6h^{-1}$ Mpc) for the \brittion{Mg}{II} absorbers. 


However, we know that the galaxies and absorbers will not occupy haloes of a single mass in this manner but will be distributed over a range of halo masses. This distribution is known as the halo occupation distribution (HOD). There are many different HODs that can yield the same value of the bias, for instance adding more high mass haloes can be balanced out with more low mass haloes. However, since the relationship between bias and halo mass is non-linear these different HODs can yield different mean halo masses for the same bias. Therefore, if we wish to estimate an accurate mean halo mass it is important to use the appropriate HOD. 

Much work has been done on determining the HOD for galaxy samples and a suitable form for luminosity limited samples is given by \citet{Zheng05} (approximately a step function for central galaxies and a powerlaw for satellites). While our galaxy sample isn't strictly luminosity limited it should yield a reasonable estimate of the mean halo mass \citep[see Appendix A of ][]{Wake08b}. If we fit using this HOD with parameters suitable for the DEEP2 galaxies and a halo model as described in \citet{Wake11} we find a higher mean halo mass of $<logM>=12.63\pm0.01\Msun$.  The same formalism may be applied to the \brittion{Mg}{II} absorbers, yielding a mean halo mass of $<logM>=12.66\pm^{0.11}_{0.15}\Msun$. 


However, there is no reason to believe that \brittion{Mg}{II} absorbers have a HOD that resembles a luminosity limited galaxy sample. Whilst the gas causing the absorption is likely to trace the underlying galaxy population its presence or absence will depend on additional physics, such as gas temperature, that will be correlated with the dark matter halo properties in a complex way (see \citet{TC08} for a \brittion{Mg}{II}-specific HOD model). With a significantly more precise measurement of the \brittion{Mg}{II} absorber-galaxy cross-correlation it will be possible to put constraints on the  \brittion{Mg}{II} absorber HOD. Such a measurement requires a larger overlapping survey, so we must reserve this analysis for future work.

One possible systematic error that could affect the bias and halo mass determinations arises if the galaxy and absorber samples have different redshift distributions and the galaxy clustering amplitude, and hence bias, depends on redshift. We show in Figure \ref{fig:1} the redshift distribution of the DEEP2 galaxies, the full DR7 \brittion{Mg}{II} absorber catalogue and the 21 \brittion{Mg}{II} absorbers in the DEEP2 area. The full DR7 \brittion{Mg}{II} sample is fairly flat with redshift and so would trace the DEEP2 sample evenly, however the \brittion{Mg}{II} absorbers in the DEEP2 area have a clear structure in their redshift distribution, which could potentially bias our calculations.

Since the DEEP2 galaxy sample is magnitude-limited, the clustering amplitude we measure could contain a redshift-dependent bias due to the preference for detecting more luminous (and thus more biased) galaxies at higher redshifts.  However, Coil et al. (2007) find no significant effect on the relevant scales and we again confirm this with our calculations.  We do see a trend on small scales ($< 1 h^{-1}$ Mpc) and some variation on the largest scales ($>20 h^{-1}$ Mpc), although this could well be caused by measurement error.

To test to what extent redshift-dependent bias affects our measurements, we select a sample of DEEP2 galaxies that match the redshift distribution of the 21 \brittion{Mg}{II} absorbers in DEEP2 area. We do this by selecting the 60 DEEP2 galaxies that have redshifts closest to the redshift of each absorber yielding a sample of 1260 galaxies. We then cross-correlate these galaxies with the full DEEP2 sample and find a cross-correlation function almost identical to the auto-correlation of the full sample (weighted mean difference $ < 1\%$ for $0.2 < s < 40$). This result indicates that the differing redshift distributions should have no effect on our results.

\section{The Galaxy Properties of Small-Scale Pairs}

\begin{figure}
\includegraphics[width=60mm, angle=270]{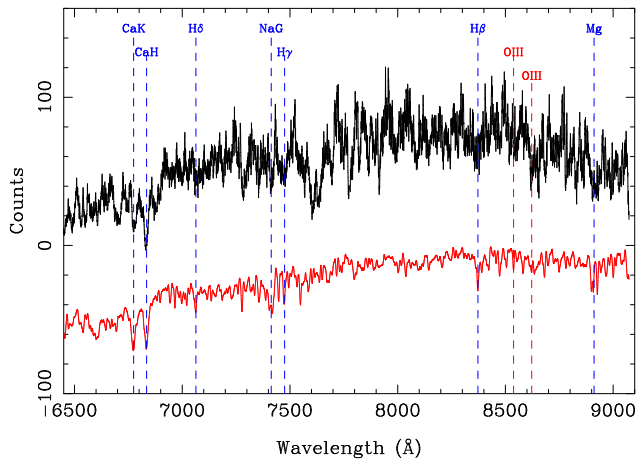}
\caption{The spectrum of galaxy DEEP2 41013534, with z=0.72218, shown in black with a 12 pixel FWHM Gaussian smoothing. The flux is uncalibrated and given in units of DEIMOS counts per hour.  The locations of specific absorption features in the galaxy are labeled in blue, and the locations of expected OIII emission are shown in red along with an example template of a 4 Gyr old single age stellar population (also shown in red). This object has the smallest projected physical separation with a \brittion{Mg}{II} absorber in our sample ($\rho_{gal}$=37 h$^{-1}$kpc), which has W$^{\lambda2796}_{r}$=1.78\AA~ and z$_{abs}$=0.72191.}
\label{fig:3}
\end{figure}

\begin{figure}
\includegraphics[width=84mm]{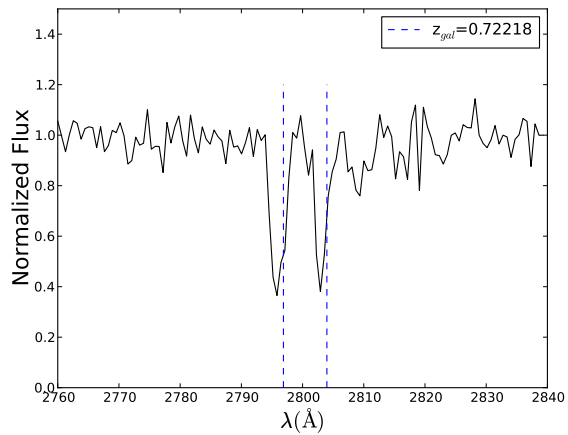}
\caption{The corresponding detection of a \brittion{Mg}{II} doublet in close proximity to the galaxy DEEP2 41013534, plotted in Figure 3.  The normalised and continuum-subtracted spectrum of quasar SDSS J022800.64+002711.6 is shown in the rest frame of detected absorption: z$_{abs}$=0.72191.  The expected location of each transition of \brittion{Mg}{II} at the redshift of the nearby galaxy (z$_{gal}$=0.72218) is overplotted with a blue dashed line.  The difference in their positions corresponds to a velocity offset of 47 km s$^{-1}$. }
\label{fig:4}
\end{figure}

Whilst the two-point cross-correlation analysis provides us with information about the distribution of \brittion{Mg}{II} absorbers relative to massive galaxies on large scales, the detection of prominent \brittion{Mg}{II} outflows in the spectra of the star-forming z$\sim$1.4 DEEP2 galaxies reported by Weiner et al. (2009) suggests that, given a sufficient number of close quasar sightlines, one might find frequent absorption coincident with these galaxies on small scales.  Adding to this expectation is the observational data indicating that L$^{*}$ galaxies typically have a cold gas covering fraction of unity on scales less than 60 h$^{-1}$kpc \citep{BB91, Bechtold92, Steideletal94}.  

Three galaxies in the DEEP2 spectroscopic sample are located within 60 h$^{-1}$kpc of quasar sightline in the SDSS DR7.  These galaxies have DEEP2 DR3 catalogue IDs of 31052516, 41013534, 41045894 and respective redshifts of 0.7346, 0.7222, and 0.8488.  The first two of the three neighboring quasar spectra exhibit \brittion{Mg}{II} absorption that is sufficiently strong so as to be identified by the automated pipeline.  These detected absorbers have redshifts of 0.7459 and 0.7219 (see ID numbers 15 and 16 in Table 1).  The third quasar spectrum has sufficient signal-to-noise such that we would expect to detect an absorber with the minimum equivalent width of our sample, should it be present.  However, we detect no \brittion{Mg}{II} absorption of this strength in the spectrum.

The \brittion{Mg}{II} absorber with z=0.7219, which has the smallest measured projected comoving separation from any DEEP2 galaxy (37 h$^{-1}$kpc), exhibits an equivalent width of 1.78\AA, which is slightly greater than the mean value for our sample (1.29\AA).  The velocity offset between the galaxy and the detected absorption is well within the 3$\sigma$ confidence intervals for the DEEP2 galaxy redshifts, making this pair a likely direct detection of both absorber and host galaxy.  The DEEP2 spectrum of the galaxy in this close pair is shown in Figure  \ref{fig:3}, and the corresponding \brittion{Mg}{II} detection is presented in Figure  \ref{fig:4}.  

While the spectroscopically identified galaxy DEEP2 41013534 has the smallest angular separation from this quasar sightline, eight other likely galaxies are identified in the DEEP2 photometric catalogue within 100 h$^{-1}$kpc of the same absorber, and three of these are detected within a projected separation of 60 h$^{-1}$kpc.  Due to the detection of more than one object within this radius, we cannot say with complete certainty that the spectroscopically observed galaxy is singly related to this \brittion{Mg}{II} absorber.  However, the nearly exact match in redshift and small projected separation makes a strong argument for a physical link between the spectroscopic galaxy and absorption complex.  

If we assume that this absorber and galaxy are associated, the physical characteristics of the galaxy spectrum can provide insight into the origin of the strong \brittion{Mg}{II} absorption complex. The spectrum of galaxy DEEP2 41013534 exhibits typical early-type characteristics.  We find no observable emission lines in this spectrum that may be attributed to active star formation.  Additionally, the rest frame U-B color of this galaxy is 1.5, placing it firmly on the red sequence.  Due to the redshift of this galaxy and the spectral coverage of the DEEP2 survey, we are unable to determine whether this galaxy exhibits evidence of intrinsic, blue-shifted \brittion{Mg}{II} absorption, which has been observed in many star-forming DEEP2 galaxies  \citep{Weiner09}. However, the absence of any other clear star-formation indicators in this galaxy implies that we cannot directly link the strong coincident \brittion{Mg}{II} absorption with a measurable rate of ongoing star formation activity in this case.  

If we consider a scenario in which the \brittion{Mg}{II} absorbing gas originated in this galaxy during an earlier period of star formation activity, we can roughly constrain the time since the original burst.  Assuming that it is emitted isotropically, gas  with an outflow velocity of 400 km s$^{-1}$, which is typical for the star-forming galaxies measured by \citet{Weiner09}, would take $\sim$100 Myr to travel the comoving distance of 37 h$^{-1}$kpc to the quasar sightline.  Stellar absorption features in the stacked spectra of similar quiescent galaxies in DEEP2 with 0.7$\la$z$\la$1 indicate that these galaxies have an average age of $\sim$1 Gyr \citep{Schiavon06}.  Considering that we are unable to constrain the specific velocity of outflows from such a star formation event in this galaxy, the estimated age of the galaxy appears to be roughly consistent with the time required for the gas to reach the separation at which it is observed.   So, while we are unable to directly connect the observed \brittion{Mg}{II} absorption in this close pair to ongoing star formation, we also cannot rule out a scenario in which the gas was emitted during an earlier period of star formation activity $\sim$1 Gyr in the past.

We emphasize that this single absorber-galaxy pair is a perilously small sample from which to draw conclusions regarding the average star-formation properties of \brittion{Mg}{II} host galaxies.  However, given the growing collection of literature linking strong \brittion{Mg}{II} absorbers with active star formation, particularly at higher redshifts, we found the contradiction presented by this absorber-galaxy pair to be at least somewhat provocative and worthy of discussion.  

\section{The \brittion{Mg}{II} Covering Fraction of DEEP2 Galaxies}

The incidence of \brittion{Mg}{II} absorption in close proximity to galaxies provides an estimate of the cold gas covering fraction, $f_{c}$, of galaxy haloes.  Measurements of $f_{c}$ determined for a range of comoving separations allow us to constrain the effective gas radius, $R_{g}$, which is critical for understanding the gas physics in galaxy evolution as well as for properly parameterizing hydrodynamic simulations.  

Among the full galaxy sample of DEEP2, 63 galaxies are detected within a projected distance of 200 h$^{-1}$kpc from a SDSS quasar sightline in which \brittion{Mg}{II} at the redshift of the galaxy would be observable outside of the Lyman-$\alpha$ forest.   If we restrict this number to quasars having observed magnitudes of m$_{i}\leq$20, the approximate SNR limit to which we can reliably detect \brittion{Mg}{II} with W$_{r}^{\lambda2796}\ga$0.6\AA~ at the median redshift of our sample, 41 galaxies remain.  For each included quasar-galaxy pair we search for \brittion{Mg}{II} in our list of 21 absorbers presented in Table 1.  Allowing for a redshift separation of $|z_{gal}-z_{abs}|<$0.008, we count the number of absorber-galaxy pairs and compare to the total number of observable quasar-galaxy pairs available in order to compute $f_{c}$ as a function of impact parameter in the rest frame of the absorbers.  The width of the redshift bin in this calculation is equivalent to the scale of the maximum outflow velocities detected in \citet{Weiner09} (1000 km s$^{-1}$) at the maximum redshift of galaxies in DEEP2, z=1.4.  


We measure the W$_{r}^{\lambda2796}\ga$0.6\AA~ \brittion{Mg}{II} covering fraction for DEEP2 galaxies to be $f_{c}$=0.5 (1 detected absorber in a sample of 2 quasar-galaxy pairs) on scales of 20--60 h$^{-1}$kpc, which decreases rapidly with increasing impact parameter.   In the range 20--100 h$^{-1}$kpc, we find a covering fraction of $f_{c}$=0.125$\pm^{0.17}_{0.103}$ (1 absorber detected among 8 possible quasar-galaxy pairs), and between 100 and 200 h$^{-1}$kpc, $f_{c}$ drops to zero (measured from a sample of 33 possible quasar-galaxy pairs).  Due to the small size of our statistical sample, we estimate the 1$\sigma$ errors on these measurements using the prescriptions of \citet{Helene84}, which assumes data with a Poisson distribution.  

It is important to emphasize that our covering fraction estimates are driven by the detection of the single absorber, which is discussed in the previous section.  Because we do not detect any galaxy-absorber pairs with larger projected separations, it is fair to speculate that the typical effective gas radius of the DEEP2 sample may be as small as $\sim$40 h$^{-1}$kpc.  The sample we examine is, however, undesirably small and limited by our conservative cuts on the SNR of the SDSS quasar spectra.  

\section{Completeness of the Galaxy and Absorber Samples on Small Scales}

The obscuration of foreground galaxies by the brighter background quasar is commonly evoked to explain the lack of detected absorber-galaxy pairs with small angular separations.  Investigating the angular incidence of DEEP2 galaxies around SDSS quasars with detected absorption, we found no galaxies within  2.8\arcsec (equivalent to a comoving distance of 30 h$^{-1}$kpc at z=1).  This finding indicates that the photometric DEEP2 catalogues are incomplete on scales smaller than this separation around the quasars we examine.  At slightly larger radii we find no evidence for photometric objects being systematically excluded for spectroscopic follow-up, so we expect our analysis to be complete beyond this radius.

In addition to this photometric incompleteness, our analysis may also be limited by the spectroscopic resolution of the data.  The galaxy redshifts in the DEEP2 catalogue have a typical 1$\sigma$ precision of $\sigma_{z}$=1.3$\times10^{-4}$ which corresponds a physical scale of $\sim$400 h$^{-1}$kpc for matter in the Hubble flow at z=1.  As such, detections of galaxies and absorbers that are physically associated at the same redshift may still have measured separations on the order of this limiting resolution.  

To investigate any possible incompleteness in the \brittion{Mg}{II} data, we re-examined the spectra of all quasars with a projected separation of $<$200 h$^{-1}$kpc from DEEP2 galaxies.  Stacking each of these 63 spectra in the rest-frame of the nearby galaxy revealed a 2$\sigma$ cumulative detection of \brittion{Mg}{II}, which grew in significance as we reduced the maximum projected separation allowed.  After examining the individual spectra contributing to the stack, we determined that no \brittion{Mg}{II} absorbers with $>$3$\sigma$ significance were undetected by the automated algorithm.  We also found that the bulk of the contribution to the \brittion{Mg}{II} absorption signal in the stacked spectrum could be attributed to the single, strong \brittion{Mg}{II} absorber with a 37 h$^{-1}$ kpc separation from a DEEP2 galaxy (ID 16 in Table 1), which has been thoroughly discussed in Section 5.   

\section{Discussion}

The bias we measure for \brittion{Mg}{II} absorbers is consistent with that of the DEEP2 galaxies, although the measurement error is substantial due to the small size of the \brittion{Mg}{II} sample.  This basic agreement suggests that strong \brittion{Mg}{II} absorbers at z$\sim$1 reside in similar environments to those of the galaxies in the DEEP2 survey.  

Furthermore, the average \brittion{Mg}{II} halo mass we estimate, 1.56$\pm$0.11$\times10^{12}\Msun$, is consistent with observations at lower redshift. \citet{L09} measured a typical halo mass of 1.8$\pm^{4.2}_{1.6}\times10^{12}h^{-1}\Msun$ for absorbers with W$_{r}^{\lambda2796}$ at z=0.6, in agreement with \citet{B06} and \citet{Gauthier09}.  The consistency with our measurement at higher redshift suggests that the halo masses of \brittion{Mg}{II} absorbers evolve very little from z=1, though the errors remain large.  

It is also worth noting that the luminosity function of z$\sim$0.65 galaxies selected by \brittion{Mg}{II} absorption in nearby quasar sightlines has been shown to peak at $M_{B}$=-20 \citep{Steideletal94}, which is consistent with the mean luminosity of the DEEP2 galaxies.  While we have not analysed a statistical sample of \brittion{Mg}{II}-selected galaxies, it seems likely that the z$\sim$1 \brittion{Mg}{II} absorbers not only trace the same environments as the DEEP2 galaxy sample but also similar types of galaxies ($\sim L_{B}^{*}$).   

\subsection{Constraining \brittion{Mg}{II} Halo Evolution}

As measurements of the bias of \brittion{Mg}{II} have never been reported at this redshift, we can provide the first constraints on theoretical models describing the evolution of absorbers and their respective haloes.  \citet{TC08} produced a model for the halo occupation distribution of cold gas at z=0.6, where the bias of \brittion{Mg}{II} has been precisely measured from cross-correlations with LRGs.  With this model, \citet{TC08} predict the probability of finding an absorber with an equivalent width W$_{r}$ in a halo of mass M$_{h}$.  In doing so, they demonstrate that the observed anti-correlation of \brittion{Mg}{II} and bias at z=0.6 may be reproduced by the absence of high density cold gas in the hot haloes of the most massive (and most biased) galaxies.  

\citet{TC10}  built on the \citet{TC08} model to incorporate the observed redshift evolution of the \brittion{Mg}{II} number density, thus enabling predictions of the absorber halo occupation distribution as a function of redshift.  The number density of \brittion{Mg}{II} absorbers in the SDSS has been shown to be roughly constant with redshift \citep{Nestor05,Prochter06,L09}.  Reconciling this non-evolving \brittion{Mg}{II} number density within the context of hierarchical growth, which produces fewer haloes at a fixed mass at higher redshifts, requires evolution in the distribution of cold gas in haloes.  One can achieve this effect by varying either the effective gas radius of the haloes or the typical absorber halo mass.  As detailed in \citet{TC10}, the observational outcomes of these two scenarios are degenerate in \brittion{Mg}{II} number density, but they diverge in predicted bias over a range in redshift.

\begin{figure}
\includegraphics[width=84mm]{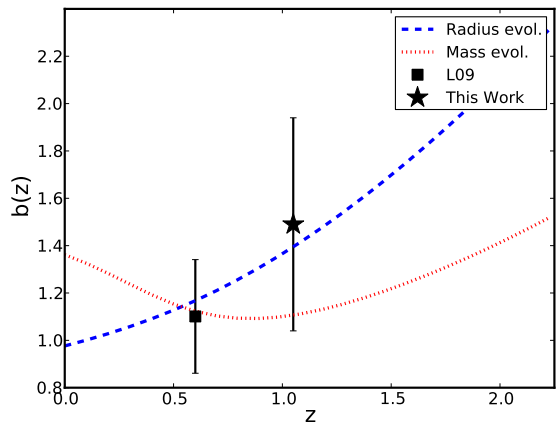}
\caption{The bias of \brittion{Mg}{II} absorbers as a function of redshift.  Model curves from Figure 3 in \citet{TC10} are overplotted, indicating the predicted bias of W$_{r}^{\lambda2796}>$1.0\AA~ absorbers for separate models of evolving halo gas radius and mass.  Bias measurements from \citet{L09} and this work are overplotted, indicating a possible preference in the data for the scenario of evolving radius.}
\label{fig:5}
\end{figure}

In Figure 5 we overlay the \brittion{Mg}{II} bias measurement from this work onto the curves of projected bias evolution calculated separately for models of evolving gas radius and halo mass from \citet{TC10}.  The bias measurement at z$\sim$1 suggests  a preference for the model of gas radius evolution, though the error on our measurement is still too large to rule out the mass evolution model.  It is important to note that the model curves of \citet{TC10} have been calculated for an absorber sample with W$_{r}>$1\AA~, whereas our data extends to a lower equivalent width limit.  However, if the anti-correlation of bias and \brittion{Mg}{II} equivalent width holds, then we would expect our bias measurement to lie slightly above the generated curve, as it does.  The amplitude of the disagreement between the model curves and our bias measurement may thus appear slightly exaggerated in this plot.  Still, any correction due to the equivalent width limit of the absorber sample is unlikely to change the measurement beyond the large, stated errors.

As detailed in \citet{TC10}, the physical implications of a model without mass evolution require the gas radius of a halo, $R_{g}$, to expand relative to the halo virial radius, $R_{200}$, with increasing redshift as
\begin{equation}
	R_{g}(M_{h},z) = R_{200}(M_{h})\times0.4\left(\frac{1+z}{1.6}\right)^{1.47}.
\label{eq:kaia}
\end{equation}
By this calculation, the gas radius should be just greater than half of the DM halo virial radius at z=1.  This implies the gas radius increases by $\sim$40\% in units of $R_{200}$ between z=0.6 and z=1.

The errors on our absorber bias measurement are largely due to the small number of \brittion{Mg}{II} absorbers available in this analysis, Thus, better constraining the absorber-galaxy cross-correlation measurement requires a deeper \brittion{Mg}{II} survey in the DEEP2 region.  The size of the \brittion{Mg}{II} sample in this work is largely determined by the SNR and resolution of SDSS spectra.  Follow-up observations of the same SDSS quasars with either higher SNR, higher resolution, or both might easily increase the number of \brittion{Mg}{II} detections, without necessitating the detection of many additional quasars within the survey footprint.   Increasing the size of the DEEP2 \brittion{Mg}{II} data  by a factor of 4, might reduce the error on the \brittion{Mg}{II} bias measurement sufficiently to rule out the \citet{Tinker10} model of evolving halo mass.

A larger dataset may also enable measurements of bias for subsets of the absorber sample split by any number of observable parameters (e,g, equivalent width).  Since previous analyses have identified a weak anti-correlation of absorber equivalent width and bias \citep{B06,L09,Gauthier09}, it would be interesting to examine whether any evolution in this trend is evident higher redshift.  Due to the limitations of our current data, we must save this potentially interesting analysis for future work.

\subsection{Comparison to Previous Measurements of the \brittion{Mg}{II} Covering Fraction}

The average effective gas radius of $\sim$ 40 h$^{-1}$kpc, estimated from our measurement of the MgII covering fraction in DEEP2 galaxies, agrees well with the speculated extent of star-formation driven outflows from the same galaxy population, which was estimated by \citet{Weiner09} to be 20-50 h$^{-1}$kpc.

Early studies found that galaxies with luminosities of $\sim L^{*}$ have a  W$_{r}^{\lambda2796}>$0.3\AA~ covering fraction of 0.25$\leq f_{c} \leq$1 on scales less than 60 h$^{-1}$kpc \citep{BB91, Bechtold92, Steideletal94}.  These analyses revealed an anti-correlation between $R_{g}$ and W$^{\lambda2796}_{r}$, which was also shown to scale slightly with galaxy luminosity.  An additional dependence of $f_{c}$ and $R_{g}$ on galaxy type was suggested by the results of \citet{Bowen95}, who used HST to examine z$<$0.2 galaxies in close proximity to quasar sightlines.   \citet{Bowen95} measured an effective gas radius $R_{g}\sim$50 h$^{-1}$kpc for galaxies with late-type or interacting morphologies and found an absence of \brittion{Mg}{II} absorption for the early-type galaxies with high confidence.

Due to the varying choices of galaxy samples, W$^{\lambda2796}_{r,min}$, and median absorber redshift, the covering fraction measurements from numerous other studies are difficult to directly compare with our results.  \citet{TB05} measured a covering fraction of $f_{c}\sim$0.5 within 60 h$^{-1}$kpc for W$_{r}^{\lambda2796}>$0.1\AA~  absorbers at z$\sim$0.5.  \citet{CT08}  measure a covering fraction of $f_{c}\sim$0.8 within 69 h$^{-1}$kpc for W$_{r}^{\lambda2796}>$0.3\AA, which seems to follow a scaling relation dependent on halo mass and galaxy luminosity.  In general agreement with this result, \citet{BC09} measure 0.25$<f_{c}<$0.4 for W$_{r}^{\lambda2796}>$0.3\AA~ within $\sim$ 75 h$^{-1}$kpc at z$\sim$0.1. 

A number of additional measurements, motivated by recent \brittion{Mg}{II}-LRG correlation results, have investigated the cold gas content of LRG haloes \citep{Chen10, Gauthier10, BC11}, which generally agree on a W$_{r}^{\lambda2796}\geq$0.6\AA~ \brittion{Mg}{II} covering fraction of $\sim$10\% for 100 h$^{-1}$kpc radii around LRGs.  However, \citet{BC11} find little evidence to support a trend of decreasing $f_{c}$ on the same scales with increasing galaxy luminosity, which might be expected if \brittion{Mg}{II} gas is evaporated at larger radii in the haloes of more luminous galaxies \citep{TC08, Chen10}.

Unfortunately, the number of absorber-galaxy pairs at separations less than 100 h$^{-1}$kpc in this work is insufficient to make a strong argument in support of any of these previous analyses.  Future surveys to obtain quasar spectra with higher resolution and SNR would be most useful in order to better constrain the gas covering fraction of these galaxies.  Such a dataset would also allow for the extension of the covering fraction measurement to lower absorber equivalent width limits.

\section{Summary}
	
We measure the two-point cross-correlation function of 21 \brittion{Mg}{II} absorbers detected in the SDSS DR7 with $\sim$32,000 spectroscopic galaxies from the DEEP2 galaxy survey in the redshift range 0.7$\leq$z$\leq$1.4.  By fitting models of the dark matter clustering to the results, we produce the first estimate of the dark matter bias and halo mass of \brittion{Mg}{II} absorbers at z$\sim$1.  

The bias we measure for the \brittion{Mg}{II} absorbers, $b_{A}$=1.49$\pm$0.45, is consistent with that of the DEEP2 galaxy sample, $b_{G}$=1.44$\pm$0.02.  This finding suggests that strong (W$_{r}^{\lambda2796}\ga$0.6\AA~) \brittion{Mg}{II} absorbers occupy similar dark matter halo environments to those of the DEEP2 galaxies.

Haloes with the bias we measure for the \brittion{Mg}{II} absorbers have a corresponding mass of 1.8$\pm^{4.2}_{1.6}\times10^{12}h^{-1}\Msun$, although the actual mean absorber halo mass will depend precisely on how these absorbers populate DM haloes.  In comparison with measurements at lower redshift, these results indicate that the dark matter halo masses of \brittion{Mg}{II} absorbers with W$_{r}^{\lambda2796}\ga$0.6\AA~ undergo no significant evolution from z$\sim$0.6. 

Since the redshift number density of these absorbers has also been shown to have roughly no evolution, these results agree with the \citet{Tinker10} model in which only the gas radius of \brittion{Mg}{II} haloes evolves with redshift.  Within this model, the gas radii of haloes expand with increasing redshift such that the haloes at z=1 have a gas radius that is $\sim$40\% larger in units of the DM halo virial radius, compared to z=0.6.  However, the errors on our \brittion{Mg}{II} bias measurement are still too large to rule out other halo evolution models in which the halo mass also evolves with redshift.

We measure the covering fraction of strong \brittion{Mg}{II} absorption to be $f_{c}$=0.5 within 60 h$^{-1}$kpc around DEEP2 galaxies.  We find no absorber host-galaxy pairs on scales larger than 37 h$^{-1}$kpc, suggesting that the effective gas radius of strong \brittion{Mg}{II} absorption around DEEP2 galaxies may be as small as $\sim$40h$^{-1}$kpc.   

In our sample, we identify just one candidate absorber host galaxy, which exhibits no evidence of ongoing star formation.  Despite the small sample size, this finding suggests that absorbers with similar equivalent widths (W$_{r}^{\lambda2796}\sim$1.8\AA) may not preferentially trace galaxies with high star formation rates at z$\sim$1.  However, we stress that a much larger sample would be required to fully test this result.

A larger overlapping survey of quasars and galaxies will be necessary to better constrain measurements of the typical environments of \brittion{Mg}{II} absorbers as well as the cold gas covering fraction of typical galaxies at z$\ga$1.  Surveys such as the SDSS-III Baryon Oscillation Spectroscopic Survey (BOSS), will soon provide the higher densities of quasars necessary to achieve this required precision.  Follow-up spectroscopy of already identified quasars at higher resolution or signal-to-noise could additionally produce the numbers of absorbers in deep galaxy survey footprints needed to vastly improve our understanding of the distribution of cold gas in dark matter haloes in the near future.



\section*{Acknowledgments}

We would like to thank the referee, Jeremy Tinker, for helpful discussions and comments, which have greatly improved this work.  We also thank JT for providing us with his model predictions, presented in Figure 5.  We are also grateful to Pushpa Khare and Jean Quashnock for helpful discussions and to many others, who have contributed to the construction of the SDSS DR7 quasar absorption line catalog of \citet{York11}.  Funding for the SDSS and SDSS-II has been provided by the Alfred P. Sloan Foundation, the U.S. Department of Energy, he National Aeronautics and Space Administration, the Japanese Monbukagakusho, the Max Planck Society, and the Higher Education Funding Council for England. The SDSS Web Site is http://www.sdss.org/.

\begin{table*}
\centering
\begin{minipage}{100mm}
\caption{\brittion{Mg}{II} Absorption Line Data}
\begin{tabular}{@{}rrrrrrrr@{}}
\hline
ID & RA   & Dec    & z$_{QSO}$ & m$_{i}$    & z$_{\brittion{Mg}{II}}$   & W$^{\lambda2796}_{r}$(\AA)  & W$^{\lambda2803}_{r}$(\AA)  \\
\hline
1 & 351.487322 & -0.083341 & 1.4082 & 18.64 & 0.9432 & 0.8301$\pm$0.1817 & 0.8980$\pm$0.2115 \\ 
2a & 351.894734 & 0.376114 & 1.4939 & 18.12 & 0.9530 & 0.7624$\pm$0.1132 & 0.5218$\pm$0.0942 \\ 
2b & .... & .... & .... & .... & 1.3378 & 0.5715$\pm$0.1120 & 0.3760$\pm$0.1035 \\ 
3 & 352.190154 & 0.083659 & 1.5272 & 19.20 & 0.9575 & 0.7678$\pm$0.1819 & 0.9195$\pm$0.2830 \\ 
4 & 353.117552 & 0.009120 & 1.6039 & 18.37 & 0.8301 & 0.9497$\pm$0.1339 & 0.4552$\pm$0.0820 \\ 
5 & 351.848711 & -0.045322 & 1.2276 & 19.85 & 0.8184 & 1.3990$\pm$0.2354 & 0.8161$\pm$0.2398 \\ 
6 & 37.391944 & 0.756865 & 1.8999 & 20.04 & 1.3018 & 2.4576$\pm$0.2789 & 1.6670$\pm$0.2828 \\ 
7 & 252.786865 & 34.942695 & 1.5404 & 18.17 & 1.2706 & 2.0955$\pm$0.1149 & 1.9330$\pm$0.1440 \\ 
8a & 252.792496 & 35.086504 & 2.2455 & 19.78 & 0.8715 & 0.8865$\pm$0.2100 & 1.0649$\pm$0.1892 \\ 
8b & .... & .... & .... & .... & 0.8765 & 1.0621$\pm$0.1886 & 0.7157$\pm$0.1913 \\ 
9 & 36.928474 & 0.667494 & 1.8943 & 19.48 & 1.3302 & 1.4213$\pm$0.2219 & 1.1321$\pm$0.2459 \\ 
10 & 36.875592 & 0.792676 & 1.4858 & 19.75 & 0.7751 & 1.9954$\pm$0.4050 & 1.0501$\pm$0.2772 \\ 
11a & 214.659800 & 52.399823 & 1.1181 & 18.68 & 0.7236 & 0.5651$\pm$0.1213 & 0.5071$\pm$0.1445 \\ 
11b & .... & .... & .... & .... & 1.0227 & 1.5198$\pm$0.1661 & 0.9206$\pm$0.1256 \\ 
12 & 215.023321 & 53.010203 & 1.6467 & 19.79 & 1.3172 & 1.3978$\pm$0.3068 & 0.7815$\pm$0.1964 \\ 
13 & 252.112461 & 35.004821 & 2.9356 & 19.08 & 1.2803 & 0.8275$\pm$0.1004 & 0.7407$\pm$0.1723 \\ 
14 & 351.815063 & 0.237062 & 2.5291 & 20.00 & 1.0394 & 2.0403$\pm$0.4246 & 3.2598$\pm$0.4487 \\ 
15 & 351.644600 & 0.363794 & 1.2571 & 20.24 & 0.7459 & 1.7217$\pm$0.3797 & 1.3855$\pm$0.3918 \\ 
16 & 37.002673 & 0.453234 & 2.3975 & 19.61 & 0.7219 & 1.7812$\pm$0.1318 & 1.3863$\pm$0.1208 \\ 
17 & 36.683492 & 0.551612 & 2.3761 & 19.90 & 1.3523 & 0.5641$\pm$0.1105 & 0.3256$\pm$0.0799 \\ 
18 & 36.790855 & 0.461854 & 2.3879 & 20.89 & 1.2268 & 1.4689$\pm$0.2991 & 0.9484$\pm$0.3121 \\ 
\hline
\end{tabular}
\end{minipage}
\end{table*}

\appendix

\bsp

\label{lastpage}

\end{document}